# Approximate scattering state solutions of DKPE and SSE with Hellmann Potential

O.J. Oluwadare[*] and K.J. Oyewumi[+]

[*]Department of Physics, Federal University Oye-Ekiti, Ekiti State, Nigeria.
[+]Theoretical Physics Section, Department of Physics, University of Ilorin, Ilorin, Nigeria.

**Abstract**

We study the approximate scattering state solutions of the Duffin-Kemmer-Petiau equation (DKPE) and the spinless Salpeter equation (SSE) with the Hellmann potential. The eigensolutions, scattering phase shifts, partial-waves transitions and the total cross section for all the partial waves are obtained and discussed. The dependence of partial-waves transitions on total angular momentum number, angular momentum number, mass combination and potential parameters were presented in the figures.

**PACS:** 03.65.-w, 03.65.Nk

**Keywords**: DKPE; SSE; Hellmann potential; scattering phase shift; partial-wave transition

## 1. Introduction

The quantum mechanical study of Hellmann potential is a time-honoured and prominent problem. Hellmann potential has already been addressed by a few number of authors (see Ref. [1-5]) since it was introduced, particularly, the bound state solutions of Hellmann potential within relativistic and non-relativistic quantum mechanics have been studied using various techniques such as Nikiforov-Uvarov method [1-2], Super-symmetric approach [3].

Just of recent, Yazarloo *et al.* extended the study to scattering states of Dirac equation with Hellmann potential under the spin and pseudospin symmetries [4]. The Dirac phase shift and normalized for the spin and pseudospin symmetries wave function were reported. Also, in one of our previous papers, we studied the scattering state solution of Klein-Gordon equation with Hellmann potential [5].

The motivation behind this work is to investigate the approximate scattering state solutions of Duffin Kemmer Petiau equation (DKPE) and spinless Salpeter equation (SSE) with Hellmann Potential. The SSE explains in detail the dynamics of semi-relativistic of particle [6-10 and the references therein] and two-body effects whereas the DKPE explains explicitly the dynamics of relativistic spin-0 and spin-1 particles [11-18]. The Hellmann potential in this study may be written as [1-5]

$$V(r) = -\frac{a}{r} + \frac{b}{r}e^{-\rho r}., \qquad (1)$$

where $a$ and $b$ are the strengths of the Coulomb potential and Yukawa potential, respectively and $\rho$ is the potential screening parameter which regulates the shape of the potential.

Section 2 presents scattering state solutions of DKPE with Hellmann potential. The scattering state solution of SSE with Hellmann potential is presented in Section 3. In Section 4, we discuss the results and the conclusion are given in Section 5.

*Corresponding author: oluwatimilehin.oluwadare@fuoye.edu.ng   [+] kjoyewumi66@unilorin.edu.ng

## 2. Scattering states of the Duffin-Kemmer-Petiau equation (DKPE) with Hellman potential

The DKP equation with energy $E_{n,J}$, total angular momentum centrifugal term and the mass m of the particle is given as [11-14]:

$$U''_{n,J}(r) - \frac{J(J+1)}{r^2} + \left[(E_{n,J} + U_v^0)^2 - m^2\right] U_{n,J}(r) = 0. \quad (2)$$

The effect of total angular momentum centrifugal term in Eq. (2) can be subdued using approximation scheme of the type [7, 10, 14]

$$\frac{1}{r^2} \approx \frac{\rho^2}{(1-e^{-\rho r})^2}. \quad (3)$$

The above approximation has been reported to be valid for $\rho r \ll 1$ [14]. By substituting Eqs. (1) and (3) into Eq. (2) and transform using mapping function $z = 1 - e^{-\rho r}$, leads to

$$z^2(1-z)^2 U''_{n,J}(z) - z^2(1-z) U'_{n,J}(z) + [-\beta_1 z^2 + \beta_2 z - \beta_3] U_{n,J}(z) = 0, \quad (4)$$

where we have employed the following parameters for simplicity

$$-\beta_1 = a\left(a - \frac{2E_{n,J}}{\rho}\right) + b\left(b - \frac{2E_{n,J}}{\rho}\right) - J(J+1) - \frac{k^2}{\rho^2}, \quad (5)$$

$$\beta_2 = -\frac{2E_{n,J}}{\rho}(a+b) + 2b(a-b), \quad (6)$$

$$-\beta_3 = J(J+1) - (a-b)^2, \quad (7)$$

and $k = \sqrt{(E_{n,J}^2 - m^2) + a^2\rho^2 - 2a\rho E_{n,J} - J(J+1)\rho^2}$ is the wave propagation constant.

Choosing the trial wave function of the type:

$$U_{n,J}(z) = z^\gamma (1-z)^{-i(k/\rho)} u_{n,J}(z), \quad (8)$$

and substituting it into Eq. (4), we obtain the hypergeometric type equation [19]

$$z(1-z) u''_{n,J}(z) + \left[2\gamma - \left(2\gamma - 2i\frac{k}{\rho} + 1\right) z\right] u'_{n,J}(z) + \left[\left(\gamma - i\frac{k}{\rho}\right)^2 + \beta_1\right] u_{n,J}(z) = 0, \quad (9)$$

where

$$\gamma = \frac{1}{2} + \sqrt{\left(J + \frac{1}{2}\right)^2 - (a-b)^2}, \quad (10)$$

$$\tau_1 = \gamma - i\frac{k}{\rho} - \sqrt{a\left(a - \frac{2E_{n,J}}{\rho}\right) + b\left(b - \frac{2E_{n,J}}{\rho}\right) - J(J+1) - \frac{k^2}{\rho^2}}, \quad (11)$$

$$\tau_2 = \gamma - i\frac{k}{\rho} + \sqrt{a\left(a - \frac{2E_{n,J}}{\rho}\right) + b\left(b - \frac{2E_{n,J}}{\rho}\right) - J(J+1) - \frac{k^2}{\rho^2}}, \quad (12)$$

$$\tau_3 = 2\gamma. \quad (13)$$

Therefore, the DKP radial wave functions for any arbitrary $J$ − states may written as:

$$U_{n,J}(r) = N_{n,J}(1 - e^{-\rho r})^\gamma e^{ikr} {}_2F_1(\tau_1, \tau_2, \tau_3; 1 - e^{-\rho r}), \quad (14)$$

where $N_{n,J}$ is the normalization factor.

The phase shifts $\delta_J$ and normalization factor $N_{n,J}$ can be obtained by applying the analytic-continuation formula [19].

$${}_2F_1(\tau_1, \tau_2, \tau_3; z) = \frac{\Gamma(\tau_3)\Gamma(\tau_3 - \tau_1 - \tau_2)}{\Gamma(\tau_3 - \tau_1)\Gamma(\tau_3 - \tau_2)} {}_2F_1(\tau_1; \tau_2; 1 + \tau_1 + \tau_2 - \tau_3; 1-z)$$

$$+ (1-z)^{\tau_3 - \tau_1 - \tau_2} \frac{\Gamma(\tau_3)\Gamma(\tau_1 + \tau_2 - \tau_3)}{\Gamma(\tau_1)\Gamma(\tau_2)} {}_2F_1(\tau_3 - \tau_1; \tau_3 - \tau_2; \tau_3 - \tau_1 - \tau_2 + 1; 1-z). \quad (15)$$

Considering Eq. (15) with the property ${}_2F_1(\tau_1, \tau_2, \tau_3; 0) = 1$, when $r \to \infty$, yields

$${}_2F_1(\tau_1, \tau_2, \tau_3; 1 - e^{-\rho r}) \xrightarrow{r \to \infty} \Gamma(\tau_3) \left|\frac{\Gamma(\tau_3 - \tau_1 - \tau_2)}{\Gamma(\tau_3 - \tau_1)\Gamma(\tau_3 - \tau_2)} + e^{-2ikr} \left|\frac{\Gamma(\tau_3 - \tau_1 - \tau_2)}{\Gamma(\tau_3 - \tau_1)\Gamma(\tau_3 - \tau_2)}\right|^*\right|, \quad (16)$$

The following relations have been introduced in the process of derivation

$$\tau_3 - \tau_1 - \tau_2 = (\tau_1 + \tau_2 - \tau_3)^* = 2i(k/\rho), \quad (17)$$

$$\tau_3 - \tau_2 = \gamma + i\frac{k}{\rho} - \sqrt{a\left(a - \frac{2E_{n,J}}{\rho}\right) + b\left(b - \frac{2E_{n,J}}{\rho}\right) - J(J+1) - \frac{k^2}{\rho^2}} = \tau_1^*, \qquad (18)$$

$$\tau_3 - \tau_1 = \gamma + i\frac{k}{\rho} + \sqrt{a\left(a - \frac{2E_{n,J}}{\rho}\right) + b\left(b - \frac{2E_{n,J}}{\rho}\right) - J(J+1) - \frac{k^2}{\rho^2}} = \tau_2^*. \qquad (19)$$

Now, defining a relation

$$\frac{\Gamma(\tau_3-\tau_1-\tau_2)}{\Gamma(\tau_3-\tau_1)\Gamma(\tau_3-\tau_2)} = \left|\frac{\Gamma(\tau_3-\tau_1-\tau_2)}{\Gamma(\tau_3-\tau_1)\Gamma(\tau_3-\tau_2)}\right| e^{i\delta}, \qquad (20)$$

and inserting it into Eq. (16) yields

$${}_2F_1(\tau_1, \tau_2, \tau_3; 1 - e^{-\rho r}) \xrightarrow{r \to \infty} \Gamma(\tau_3)\left[\frac{\Gamma(\tau_3-\tau_1-\tau_2)}{\Gamma(\tau_3-\tau_1)\Gamma(\tau_3-\tau_2)}\right] e^{-ikr}\left[e^{i(kr-\delta)} + e^{-i(kr-\delta)}\right] \qquad (21)$$

Thus, we have the asymptotic form of Eq. (14) when $r \to \infty$ as;

$$U_{n,J}(r) \xrightarrow{r \to \infty} 2N_{n,J}\Gamma(\tau_3)\left[\frac{\Gamma(\tau_3-\tau_1-\tau_2)}{\Gamma(\tau_3-\tau_1)\Gamma(\tau_3-\tau_2)}\right] \sin\left(kr + \frac{\pi}{2} + \delta\right). \qquad (22)$$

Accordingly, with the appropriate boundary condition imposed by Ref. [20], Eq. (22) yields

$$U_{n,J}(\infty) \to 2\sin\left(kr - \frac{l\pi}{2} + \delta_J\right). \qquad (23)$$

and on comparison of Eq. (22), the DKP phase shift and the corresponding normalization factor can be found respectively as:

$$\delta_J = \frac{\pi}{2} + \frac{J\pi}{2} + \delta = \frac{\pi}{2}(J+1) + arg\Gamma(2ik/\rho) - arg\Gamma(\tau_2^*) - arg\Gamma(\tau_1^*) \qquad (24)$$

and

$$N_{n,J} = \frac{1}{\sqrt{\tau_3}}\left|\frac{\Gamma(\tau_1^*)\Gamma(\tau_2^*)}{\Gamma(2i(k/\rho))}\right|. \qquad (25)$$

The DKP total cross-section for the sum of partial-wave cross-sections $\sigma_l$ is defined as [10]:

$$\sigma_{total} = \sum_{l=0}^{\infty} \sigma_l = \frac{4\pi}{k^2}\sum_{l=0}^{\infty}(2l+1)\sin^2\delta_l \qquad (26)$$

where $T_J = 4\sin^2\delta_J$ defines the DKP partial-wave transitions.

A straightforward substitution of Eq. (24) into Eq. (26) yields the total cross-section

$$\sigma_{total} = \frac{4\pi}{k^2}\sum_{l=0}^{\infty}(2l+1)\sin^2\left[\frac{\pi}{2}(J+1) + arg\Gamma(2ik/\rho) - arg\Gamma(\tau_2^*) - arg\Gamma(\tau_1^*)\right] \qquad (27)$$

Also, we need to analyze the gamma function $\Gamma(\tau_3 - \tau_1)$ [20] from the S-matrix as:

$$\tau_3 - \tau_1 = \gamma + i\frac{k}{\rho} + \sqrt{a\left(a - \frac{2E_{n,J}}{\rho}\right) + b\left(b - \frac{2E_{n,J}}{\rho}\right) - J(J+1) - \frac{k^2}{\rho^2}}. \qquad (28)$$

The first order poles of $\Gamma\left(\gamma + i\frac{k}{\rho} + \sqrt{a\left(a - \frac{2E_{n,J}}{\rho}\right) + b\left(b - \frac{2E_{n,J}}{\rho}\right) - J(J+1) - \frac{k^2}{\rho^2}}\right)$ are situated at

$$\Gamma\left(\gamma + i\frac{k}{\rho} + \sqrt{a\left(a - \frac{2E_{n,J}}{\rho}\right) + b\left(b - \frac{2E_{n,J}}{\rho}\right) - J(J+1) - \frac{k^2}{\rho^2}}\right) + n = 0 \; (n = 0, 1, 2, \ldots). \qquad (29)$$

By applying algebraic means to Eq. (29), we obtain the DKP bound state energy levels equation for the Hellman potential as:

$$k^2 = -\rho^2\left[\frac{(n+\gamma)^2 + a\left(\frac{2E_{n,J}}{\rho} - a\right) + b\left(\frac{2E_{n,J}}{\rho} - b\right) - J(J+1)}{2(n+\gamma)}\right]^2. \qquad (30)$$

## 2. Scattering states solutions of the spinless Salpeter equation (SSE) with Hellmann potential

The spinless Salpeter equation for two different particles interacting in a spherically symmetric potential in the center of mass system is given by [21-24]:

$$\left[\sum_{i=1,2}\left(\sqrt{-\Delta + m_i^2} - m_i\right) + (V(r) - E_{n,l})\right]\chi(r) = 0, \tag{31}$$

where $\chi(r) = R_{nl}(r)Y_{lm}(\theta, \varphi)$. Also, using appropriate transformation equation $R_{nl}(r) = \psi_{nl}(r)/r$, the radial component of SSE in the case of heavy interacting particles may be written as [See details in Ref. 21-25]

$$\psi''_{nl}(r) + \left[-\frac{l(l+1)}{r^2} + 2\mu(E_{n,l} - V(r)) + \left(\frac{\mu}{\eta}\right)^3(E_{n,l} - V(r))^2\right]\psi_{nl}(r) = 0, \tag{32}$$

where

$$\mu = \frac{m_1 m_2}{(m_1 + m_2)}, \tag{33}$$

$$\left(\frac{\eta}{\mu}\right)^3 = \frac{m_1 m_2}{(m_1 m_2 - 3\mu^2)}. \tag{34}$$

The units $\hbar = c = 1$ have been employed in the process of derivation and $E_{n,l}$ is the semi-relativistic energy of the two particles having arbitrary masses $m_1$ and $m_2$. $\mu$ and $\eta$ are the reduced mass and mass index respectively. The solution to Eq. (32) becomes non-relativistic as the term with the mass index tends to zero.

Substituting the potential in Eq. (1) and approximation in Eq. (3) into Eq. (32) and applying the same procedure in section 2, the radial wave functions for the spinless Salpeter equation with Hellmann potential as:

$$\psi_{nl}(r) = N_{n,l}(1 - e^{-\rho r})^v e^{ikr} {}_2F_1(\xi_1, \xi_2, \xi_3; 1 - e^{-\rho r}). \tag{35}$$

having the following useful parameters:

$$k = \sqrt{2\mu(E_{n,l} + a\rho) + (\mu/\eta)^3(E_{n,l} + a)^2 - l(l+1)\rho^2} \tag{36}$$

$$v = 1/2 + \sqrt{(l + 1/2)^2 - (\mu/\eta)^3(a/\rho - b)^2}, \tag{37}$$

$$\xi_1 = v - i\frac{k}{\rho} - \sqrt{\frac{2\mu a}{\rho} + (\mu/\eta)^3\left[\frac{2E_{n,l}}{\rho}\left(\frac{a}{\rho} - b\right) + \left(\frac{a}{\rho} - b\right)\left(\frac{a}{\rho} + b\right)\right] - l(l+1) - \frac{k^2}{\rho^2}}, \tag{38}$$

$$\xi_2 = v - i\frac{k}{\rho} + \sqrt{\frac{2\mu a}{\rho} + (\mu/\eta)^3\left[\frac{2E_{n,l}}{\rho}\left(\frac{a}{\rho} - b\right) + \left(\frac{a}{\rho} - b\right)\left(\frac{a}{\rho} + b\right)\right] - l(l+1) - \frac{k^2}{\rho^2}}, \tag{39}$$

$$\xi_3 = 2v. \tag{40}$$

The corresponding phase shift for the spinless Salpeter equation containing Hellman potential is obtained as

$$\delta_l = \frac{\pi}{2}(l + 1) + arg\Gamma(2i(k/\rho)) - arg\Gamma(\xi_2^*) - arg\Gamma(\xi_1^*), \tag{41}$$

$$\xi_1^* = \xi_3 - \xi_2 = v + i\frac{k}{\rho} - \sqrt{\frac{2\mu a}{\rho} + (\mu/\eta)^3\left[\frac{2E_{n,l}}{\rho}\left(\frac{a}{\rho} - b\right) + \left(\frac{a}{\rho} - b\right)\left(\frac{a}{\rho} + b\right)\right] - l(l+1) - \frac{k^2}{\rho^2}}, \tag{42}$$

$$\xi_2^* = \xi_3 - \xi_1 = v + i\frac{k}{\rho} + \sqrt{\frac{2\mu a}{\rho} + (\mu/\eta)^3\left[\frac{2E_{n,l}}{\rho}\left(\frac{a}{\rho} - b\right) + \left(\frac{a}{\rho} - b\right)\left(\frac{a}{\rho} + b\right)\right] - l(l+1) - \frac{k^2}{\rho^2}}. \tag{43}$$

and the normalization constant

$$N_{n,l} = \frac{1}{\Gamma(\xi_3)}\left|\frac{\Gamma(\xi_1^*)\Gamma(\xi_2^*)}{\Gamma(2i(k/\rho))}\right|. \tag{44}$$

The energy eigenvalue equation for the spinless Salpeter equation with Hellmann potential as:

$$k^2 = -\rho^2 \left[ \frac{(n+v)^2 - \frac{2\mu a}{\rho} + (\mu/\eta)^3 \left( \frac{2bE_{n,l}}{\rho} - \frac{2aE_{n,l}}{\rho^2} - \frac{a^2}{\rho^2} + b^2 \right) + l(l+1)}{2(n+v)} \right]^2. \quad (45)$$

The total scattering cross-section for the sum of partial-wave cross-sections $\sigma_l$ is given as

$$\sigma_{tot.} = \sum_{l=0}^{\infty} \sigma_l = \frac{\pi}{k^2} \sum_{l=0}^{\infty} (2l+1) \, T_l, \quad (46)$$

where

$$T_l = 4\sin^2 \delta_l \quad (47)$$

which defines the partial-wave transitions for the SSE with Hellmann potential in this present study.

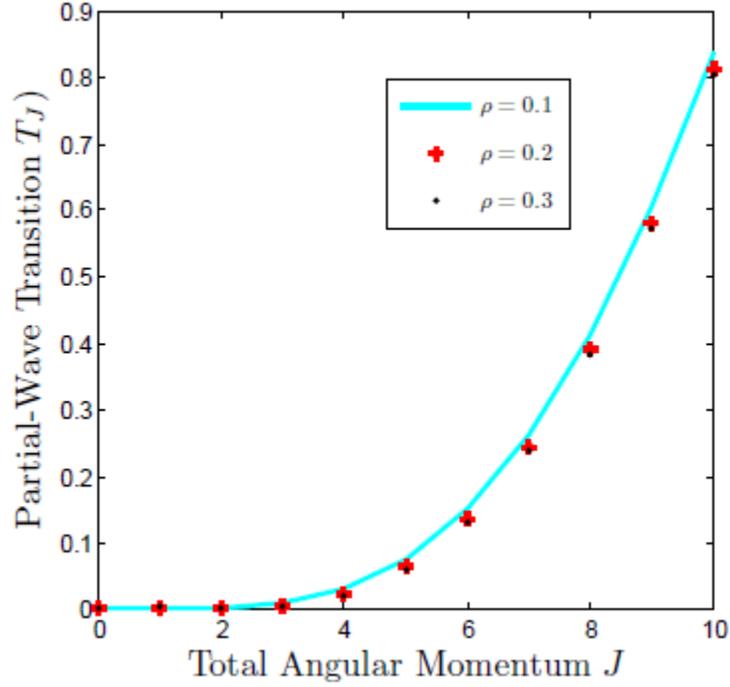

**Figure 1**: DKP Partial-wave transition for the Hellman potential as a function of total angular momentum $J$ with $a = b = 0.15$ and $E_{n,J} = m = 1$.

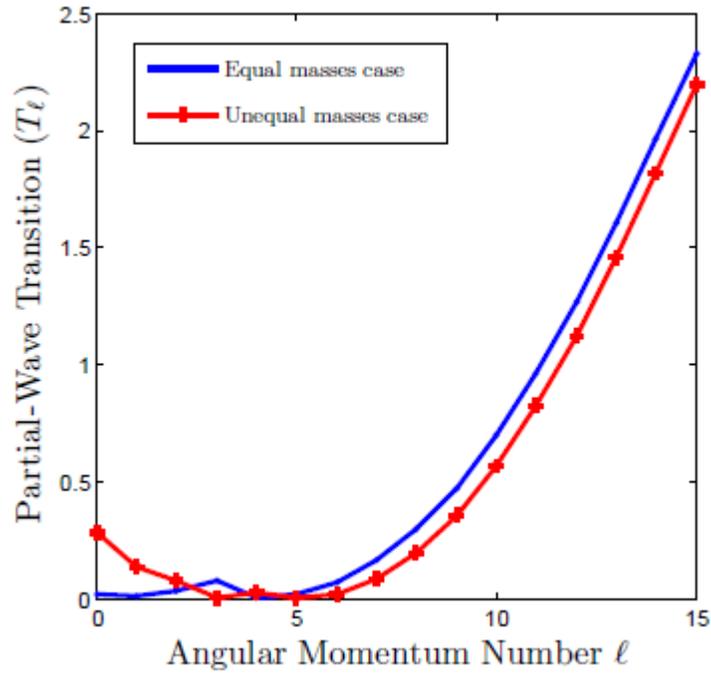

Figure 2a: Partial-wave transition for the spinless Salpeter equation with the Hellmann potential as a function of angular momentum quantum number $l$ with $a = 0.2, b = -1, E_{n,l} = 1, \rho = 0.5$.

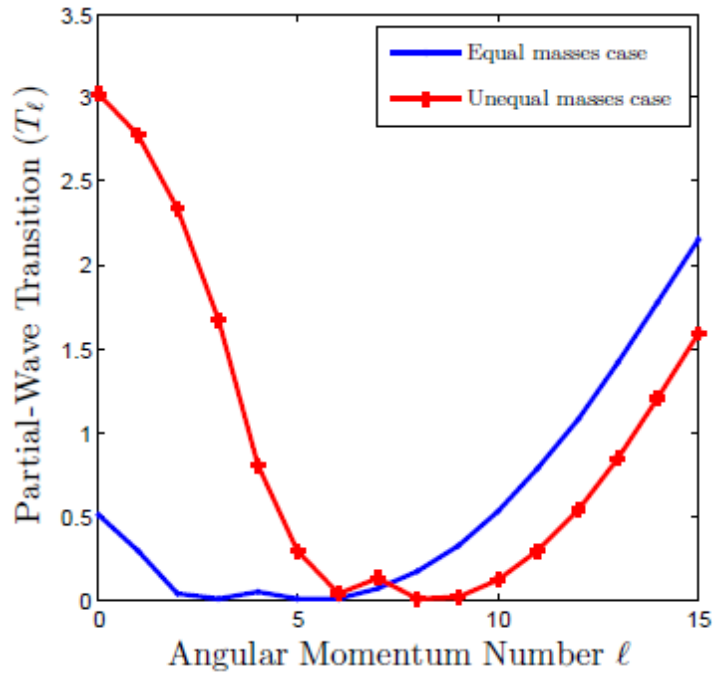

Figure 2b: Partial-wave transition for the spinless Salpeter equation with the Hellmann potential as a function of angular momentum quantum number $l$ with $a = 2, b = -1, E_{n,l} = 1, \rho = 0.5$.

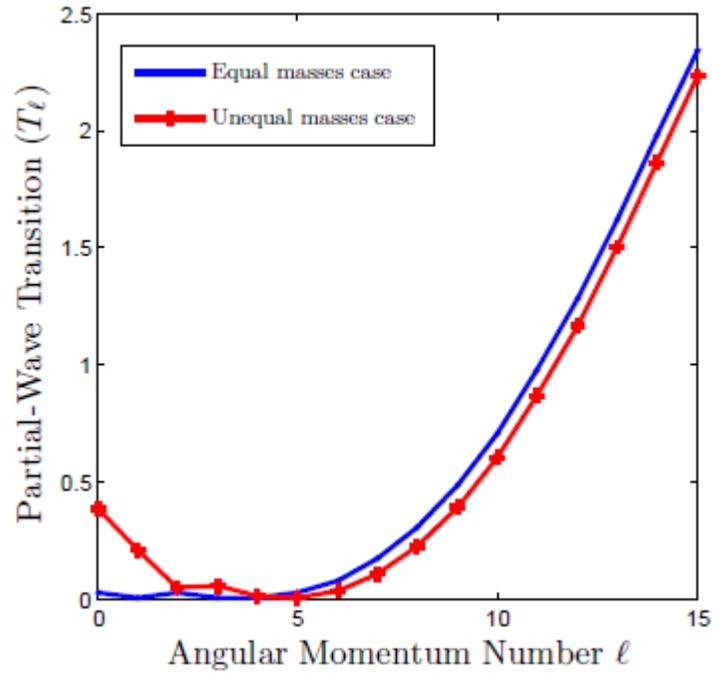

Figure 3: Partial-wave transition for the spinless Salpeter equation with the Hellmann potential as a function of angular momentum quantum number $l$ with $a = 0$, $b = -3$, $E_{n,l} = 1$, $\rho = 0.5$.

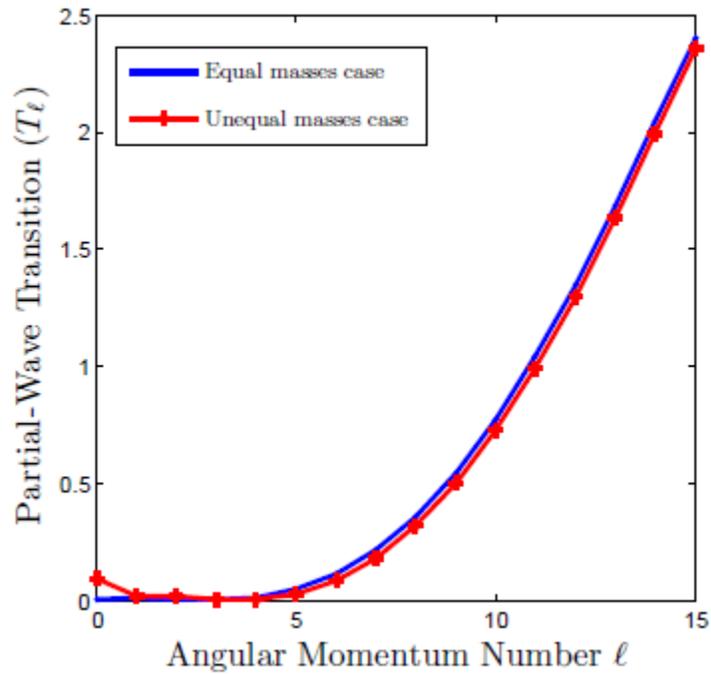

Figure 4a: Partial-wave transition for the spinless Salpeter equation with the Hellmann potential as a function of angular momentum quantum number $l$ with $a = -2$, $b = 0$, $E_{n,l} = 1$, $\rho = 0.5$.

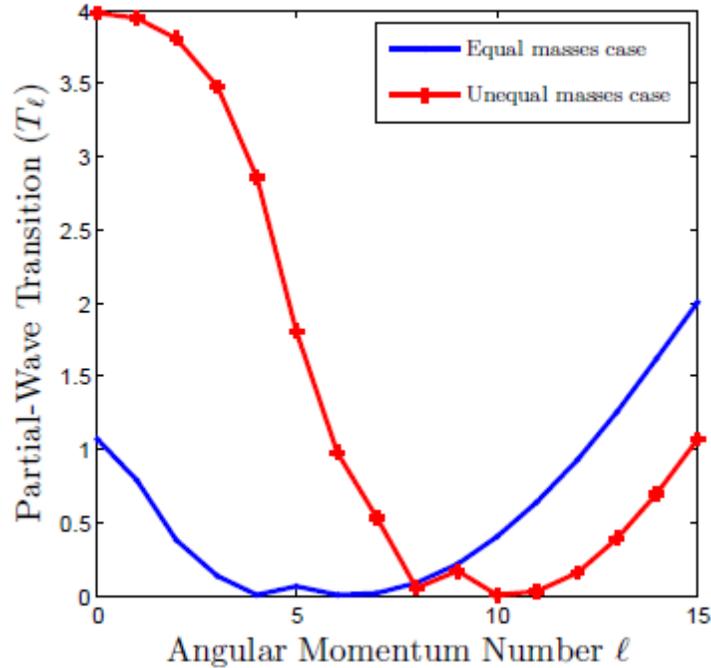

Figure 4b: Partial-wave transition for the spinless Salpeter equation with the Hellmann potential as a function of angular momentum quantum number $l$ with $a = 3, b = 0, E_{n,l} = 1, \rho = 0.5$.

## 4. Discussion

We have used the units $\hbar = c = 1$ in partial-wave transition illustrations. For equal mass cases, we used $(\mu/\eta)^3 = 1/4$ and $\mu = m_1/2$ while $(\mu/\eta)^3 = 1$ and $\mu = \frac{m_1}{100}$ were used for unequal masses cases. In all the cases, we consider $m_2 = E_{n,l} = 1$ and $m_1 = 1$ for the equal masses case only. For the screening parameters $\rho = 0.1$, $\rho = 0.2$ and $\rho = 0.3$ the DKP partial-waves transitions increases exponentially. See Figure 1. The two-body effect here appears as a shift of the phases of the partial waves. For lower partial-waves, says $l < 5$, the partial-waves transition decay exponentially whereas for higher partial waves, says $l > 5$, the partial-waves transition rise exponentially. See Figures 2-4. Also alteration of potential parameters has a serious effect on the partial-wave transition illustrations. Compare Figure 2a with 2b & 3, and Figures 4a with 4b.

## 4. Conclusion

We have investigated the approximate scattering state solutions of DKPE and SSE with Hellman potential via analytical method. The approximate DKP and semi-relativistic scattering phase shifts, partial-wave transitions, eigenvalues and normalized eigenfunctions have been obtained. The DKP and semi-relativistic partial wave transition calculations for the Hellmann potential were shown in the Figures 1 and Figures 2-4 respectively.

It is clearly shown both the total angular momentum number, angular momentum number and potential parameters contribute significantly to the partial wave transition and that the two-body effects modify the phases of the partial-waves and is usually noticeable for lower partial-waves.

**Conflict of Interest**

The authors declare that there is no conflict of interest regarding this paper.